\documentclass[10pt,aps,prd,superscriptaddress,nofootinbib,nobibnotes,longbibliography,floatfix,twocolumn]{revtex4-2}

\usepackage{bm}
\usepackage{mathtools,
amsmath,
amssymb,
amsfonts,
mathrsfs,
chngcntr,
multirow}

\let\cc\corresponds
\let\corresponds\relax
\usepackage{mathabx}
\let\corresponds\cc
\usepackage[utf8]{inputenc}
\usepackage[T1]{fontenc}

\usepackage[dvipsnames]{xcolor}
\usepackage[unicode]{hyperref}
\hypersetup{colorlinks=true, citecolor=MidnightBlue,
            linkcolor=MidnightBlue, urlcolor=MidnightBlue, linktocpage=true}
\usepackage[normalem]{ulem}

\usepackage{graphicx}
\usepackage{verbatim}

\newcommand{\dd}{\mathrm{d}}

\begin{document}
\pdfoutput=1
\title{
Strong breaking of black-hole uniqueness from coexisting scalarization mechanisms
}

\author{Astrid Eichhorn}
\email{eichhorn@thphys.uni-heidelberg.de}
\affiliation{Institut f\"ur Theoretische Physik, Universit\"at Heidelberg, Philosophenweg 12 \& 16, 69120 Heidelberg, Germany}

\author{Pedro G. S. Fernandes}
\email{fernandes@thphys.uni-heidelberg.de}
\affiliation{Institut f\"ur Theoretische Physik, Universit\"at Heidelberg, Philosophenweg 12 \& 16, 69120 Heidelberg, Germany}

\author{Lidia Marino}
\email{marino_l@thphys.uni-heidelberg.de}
\affiliation{Institut f\"ur Theoretische Physik, Universit\"at Heidelberg, Philosophenweg 12 \& 16, 69120 Heidelberg, Germany}

\begin{abstract}
Black-hole uniqueness, i.e., the statement that all stationary vacuum black holes in the universe are described by the Kerr solution, is expected to break in theories beyond General Relativity. This breaking can take a particularly strong form, if several branches of black-hole solutions beyond the Kerr solution coexist. We find an example of a theory that exhibits such strong breaking. In this theory, a cubic coupling of a scalar field to the Gauss-Bonnet invariant triggers black-hole scalarization through a non-linear instability of the Kerr solution. At large spin, curvature-induced and spin-induced scalarization mechanisms compete at fixed sign of the coupling. This results in a rich phase structure of black-hole solutions and continuous as well as discontinuous transitions between the different branches of black holes.
\end{abstract}

\maketitle

\section{Introduction}
The black-hole uniqueness theorems~\cite{PhysRevLett.26.331,PhysRevLett.34.905} of General Relativity (GR) make one of its most striking predictions, namely that black holes across all masses are described by the Kerr solution~\cite{PhysRevLett.11.237}, in which only the mass and spin are free parameters. However, astrophysical black holes cannot be fully described by the Kerr solution of General Relativity (GR). While the spacetime outside of the event horizon is free of obvious problems, the interior of a Kerr black hole is not. Besides the curvature singularity, at which tidal forces blow up and at which geodesics terminate in finite proper time, there is a Cauchy horizon at which the initial-value problem breaks down and there are closed timelike curves. The Cauchy horizon typically lies in a region of spacetime with low curvature (compared to the Planck scale). Its existence is the strongest theoretical indication we have for new physics in gravity at curvature scales far below the Planck scale.

By continuity arguments, a modified, potentially quantum theory of gravity, which modifies the black-hole interior and solves the above problems, is expected to not leave the black-hole exterior completely unmodified\footnote{For a counterexample to this expectation, see the gravastar solution in \cite{Eichhorn:2025pgy}.}. Thus, one may expect that as observations of black holes, through gravitational waves \cite{LIGOScientific:2016aoc} and black-hole imaging \cite{EventHorizonTelescope:2019dse,EventHorizonTelescope:2022wkp}, continue to improve in sensitivity, the predictions of GR and the observational data may at some point no longer agree. The sensitivity required to achieve this may be extremely high, if the scale at which modifications from GR set in is the Planck scale, as generally expected from quantum gravity.\footnote{See, however, \cite{Eichhorn:2022bbn,Horowitz:2023xyl} for arguments how quantum-gravity effects are enhanced at near-extremal values of the spin.} However, it may well be that GR is modified at much lower curvature scales. Hints for a modification of GR far from the Planck scale come from cosmological observations, where different measurements of the $\Lambda$CDM parameters are in tension, see e.g.~\cite{DiValentino:2021izs}, and tentative hints at dynamics beyond $\Lambda$CDM exist \cite{DESI:2024mwx,DESI:2025zgx}, although these results are not yet robust enough to jump to strong conclusions. In addition, Starobinsky-type inflation \cite{Starobinsky:1980te}, requires a curvature-squared coupling many orders of magnitude larger than the natural expectation, continues to fit observational data \cite{Addazi:2025qra} and hint at a new-physics scale in gravity that is decidedly lower than the Planck scale.

In order to search for deviations from GR in observations of black holes, it is necessary to map out -- at least in part -- the space of possible modifications. Beyond GR, it is expected that black-hole uniqueness no longer holds and different (possibly stable) branches of black-hole solutions may exist at a given value of mass and spin. Which of these solutions is realized for any given astrophysical black hole may then, for example, depend on its formation history. Then, different populations of black holes may well be described by different solutions that might, in addition to mass and spin, be characterized by additional parameters colloquially referred to as \emph{hair}. The potential of future observatories to deliver a comprehensive view of black-hole populations across a large range of redshifts \cite{Arca-Sedda:2025vhw} may thus be crucial for the understanding of how GR should be modified.

One class of theories in which black-hole uniqueness is known to be violated is scalar–Gauss–Bonnet gravity. In these models, suitable couplings between a scalar field and the Gauss–Bonnet invariant can under appropriate conditions cause the scalar to acquire an effective, spacetime-dependent tachyonic mass near the horizon of a Kerr black hole. 
This triggers a tachyonic instability whose endpoint is a stationary black-hole solution that deviates from Kerr. This process is referred to as black-hole \emph{scalarization} \cite{Doneva:2017bvd,Silva:2017uqg}, see \cite{Doneva:2022ewd} for a review, and~\cite{Doneva:2017bvd,Silva:2017uqg,Cunha:2019dwb,Dima:2020yac,Herdeiro:2020wei,Berti:2020kgk,Doneva:2022ewd,Doneva:2021tvn,Doneva:2022yqu,Staykov:2022uwq,Lai:2023gwe,Doneva:2023kkz,Pombo:2023lxg,Belkhadria:2023ooc,Xiong:2023bpl,Zhang:2024spn,Liu:2025eve,Herdeiro:2018wub,Witek:2018dmd,Silva:2018qhn,Blazquez-Salcedo:2018jnn,Fernandes:2019rez,Macedo:2019sem,East:2021bqk,Minamitsuji:2018xde,Silva:2020omi,Collodel:2019kkx,Doneva:2018rou,Konoplya:2019fpy,Andreou:2019ikc,Corman:2022xqg,Doneva:2020nbb,Doneva:2017duq,Doneva:2019vuh,Fernandes:2019kmh,Hod:2020jjy,Ripley:2020vpk,Elley:2022ept,Blazquez-Salcedo:2020rhf,Antoniou:2021zoy,Witek:2020uzz,Herdeiro:2019yjy,Fernandes:2020gay,Kuan:2021lol,Hod:2019pmb,Ramazanoglu:2019gbz,Doneva:2021dqn,Zhang:2021nnn,Anson:2019uto,Antoniou:2020nax,Doneva:2022byd,Ventagli:2020rnx,Doneva:2020kfv,Antoniou:2022agj,Doneva:2023oww,Zhang:2022cmu,Blazquez-Salcedo:2022omw,Liu:2022fxy,Thaalba:2023fmq,Danchev:2021tew,Herdeiro:2021vjo,Liu:2020yqa,Kuan:2023trn,Ventagli:2021ubn,Anson:2019ebp,Eichhorn:2023iab,Lara:2024rwa,Doneva:2020qww,Thaalba:2025ljh,Doneva:2024ntw,Staykov:2025lfh,Bahamonde:2022chq,Fernandes:2022kvg,Minamitsuji:2023uyb,Fernandes:2024ztk,AresteSalo:2025sxc,Annulli:2022ivr,Thaalba:2024htc,Liu:2025bkz,Staykov:2021dcj,Xiong:2024urw,Eichhorn:2025aja,Belkhadria:2025lev,AresteSalo:2026zzc,Ye:2026nvr,Muniz:2025egq} for a comprehensive list of works in the field.

Scalarization can occur in two distinct ways. The first is \emph{curvature-induced} scalarization \cite{Doneva:2017bvd,Silva:2017uqg,Cunha:2019dwb}, in which the instability is triggered by high curvature, as encoded in the Gauss-Bonnet invariant. This is realized near the horizon of a Kerr black hole at sufficiently small mass (in units of the Gauss-Bonnet coupling), independently of the spin.
The second is \emph{spin-induced} scalarization \cite{Dima:2020yac,Berti:2020kgk,Herdeiro:2020wei}, in which the instability is triggered by the curvature becoming  \emph{negative} and large in absolute value near the horizon of a highly spinning Kerr black hole.

Spin- and curvature-induced scalarization mechanisms hinge on the scalar-Gauss-Bonnet interaction contributing negatively to the potential of the scalar field evaluated on the background of a Kerr black hole.
If the scalar field is $\mathbb{Z}_2$ symmetric, as has been assumed in most of the existing literature, the sign of the coupling between the scalar field and the Gauss-Bonnet invariant determines whether spin-induced or curvature-induced scalarization is feasible.
A given theory with a $\mathbb{Z}_2$ symmetric scalar field admits only one type of scalarization. As a result, the two mechanisms cannot coexist within the same theory.

In this work, we present the first example of a scalar–Gauss–Bonnet theory, in which both spin- and curvature-induced scalarization mechanisms coexist within certain regions of the parameter space. This coexistence leads to a \emph{strong} violation of black-hole uniqueness: in addition to the Kerr solution, two distinct scalarized black-hole solutions are admitted. 

In our example, the scalarization mechanism does not arise from a linear tachyonic instability, but from non-linear effects, see, e.g., Refs.~\cite{Doneva:2021tvn,Doneva:2022yqu,Staykov:2022uwq,Lai:2023gwe,Doneva:2023kkz,Pombo:2023lxg,Belkhadria:2023ooc,Xiong:2023bpl,Zhang:2024spn,Liu:2025eve} for other examples of non-linear scalarization. This non-linear effect, within a setting without $\mathbb{Z}_2$ symmetry, is causing the \emph{strong} violation of black-hole uniqueness.

This paper is organized as follows: In Section \ref{Our theory} we introduce the theory and discuss how several scalarization mechanisms can coexist; in Section \ref{Methodology} we discuss the methodology adopted to solve the field equations. Section \ref{Results} presents the bulk of our results including the phase diagram for black-hole solutions. We conclude in Section \ref{Conclusions}, and discuss future directions.

\section{scalar-Gauss-Bonnet gravity with a cubic coupling} \label{Our theory}
We consider a scalar-Gauss-Bonnet theory of gravity based on the Gauss-Bonnet invariant $\mathcal{G}$, non-minimally coupled to a scalar field $\phi$
\begin{equation}
    S = \frac{1}{16\pi} \int d^4 x \sqrt{-g} \Bigg( R - \nabla_{\mu} \phi\nabla^{\mu} \phi + \frac{\alpha}{4} f(\phi) 
    \mathcal{G} \Bigg).
    \label{eq:action}
\end{equation} 
We take the coupling function to be cubic
\begin{equation}
    f(\phi) = \frac{\phi^3}{6}.
    \label{eq:coupling}
\end{equation}
Generally, the Taylor expansion of the coupling function determines the solutions: if $f(\phi) \sim \mathcal{O}(\phi^2)$, the scalar field can remain trivial on top of a vacuum solution of the Einstein equations.
Consequently, the Kerr metric is a solution of our theory, with $\phi=0$. The question we address in this section is whether the Kerr metric is the unique stationary and axially-symmetric black-hole solution, or not. When the coupling function is quadratic, under appropriate conditions, the Kerr metric can become linearly unstable and scalarize~\cite{Doneva:2017bvd,Silva:2017uqg,Cunha:2019dwb,Dima:2020yac,Berti:2020kgk,Herdeiro:2020wei}. This instability is a linear tachyonic instability. This follows from the equation of motion for the scalar field
\begin{equation}
    \Box \phi = -\frac{\alpha}{8} f'(\phi) \mathcal{G},
\end{equation}
where the combination $-\frac{\alpha}{8} f''(0) \mathcal{G}$ acts an effective mass squared for scalar perturbations, which can become negative.

For a cubic coupling function~\eqref{eq:coupling}, however, there is no such instability, as the effective mass squared of scalar perturbations always vanishes. Nonetheless, non-linear effects could trigger an instability and give rise to additional branches of black-hole solutions~\cite{Doneva:2021tvn}.

To understand under which conditions additional branches of solutions might exist, we examine the scalar field equation of motion, that for the coupling~\eqref{eq:coupling} becomes
\begin{equation} \label{eq:EoM scalar field}
    \Box \phi + \frac{\alpha}{16} \phi^2 \, \mathcal{G} = 0.
\end{equation}
Following Ref.~\cite{Herdeiro:2015waa}, we examine when the conditions for a no-hair theorem, using only the equation of motion for the scalar field, fail. Assuming the scalar field inherits the symmetries of the spacetime (stationarity and axial-symmetry), we multiply the scalar field equation by $\phi$, and integrate over the exterior black-hole spacetime. Integrating by parts, we find
\begin{equation}
\int d^4x \sqrt{-g} \, \nabla_\mu \phi \, \nabla^\mu \phi = \frac{\alpha}{16} \int d^4x \sqrt{-g} \, \phi^3 \mathcal{G}.\label{eq:integratedeom}
\end{equation}
Due to the Killing symmetries, the gradient of the scalar field may only depend on the non-Killing coordinates and thus the gradient is spacelike (or vanishes). As a consequence, in $(-,+,+,+)$-signature, the left-hand-side of Eq.~\eqref{eq:integratedeom} is always positive.\footnote{For the other sign convention for the metric, the same argument goes through with opposite signs for the coupling $\alpha$.} Therefore, the following inequality must always be respected:
\begin{equation} \label{conditionSF1}
\alpha \int d^4x \sqrt{-g} \, \phi^3 \mathcal{G} \geq 0.
\end{equation}
This inequality is trivially satisfied for a trivial scalar solution, but it is not strong enough to exclude non-trivial solutions. Instead, it simply constrains the combination of sign of the coupling, Gauss-Bonnet and scalar field.

In Boyer-Lindquist coordinates $(t, \rho, \theta, \varphi)$, the Gauss-Bonnet invariant of the Kerr metric is given by
\begin{equation}
    \begin{aligned}
        \mathcal{G}(\rho, \theta) =& \frac{48 M^2}{(\rho^2 + j^2 M^2 \cos^2 \theta)^6}  \Big( \rho^6 - 15 \rho^4 j^2 M^2 \cos^2 \theta\\&
        + 15 \rho^2 j^4 M^4 \cos^4 \theta - j^6 M^6 \cos^6 \theta \Big).
    \end{aligned}
\end{equation}
Focusing on the region outside of the horizon, located at $\rho_H= M\left(1 + \sqrt{1- j^2}\right)$, for small spins, the Gauss-Bonnet invariant is positive definite in this region.
In contrast, when $j\geq1/2$, regions with negative $\mathcal{G}$ exist outside the horizon, starting at the poles. These regions grow as the spin increases, until in the extremal case, $j=1$, there is a finite region with negative $\mathcal{G}$ and a finite region with positive $\mathcal{G}$.
Thus, assuming $\phi\neq 0$, we can 
identify two distinct  
ways in
which the no-hair theorem is violated (i.e., the inequality \eqref{conditionSF1} is satisfied non-trivially):

\begin{figure}[!t]
    \includegraphics[width=\linewidth,clip=true, trim=3cm 0cm 19cm 0cm]{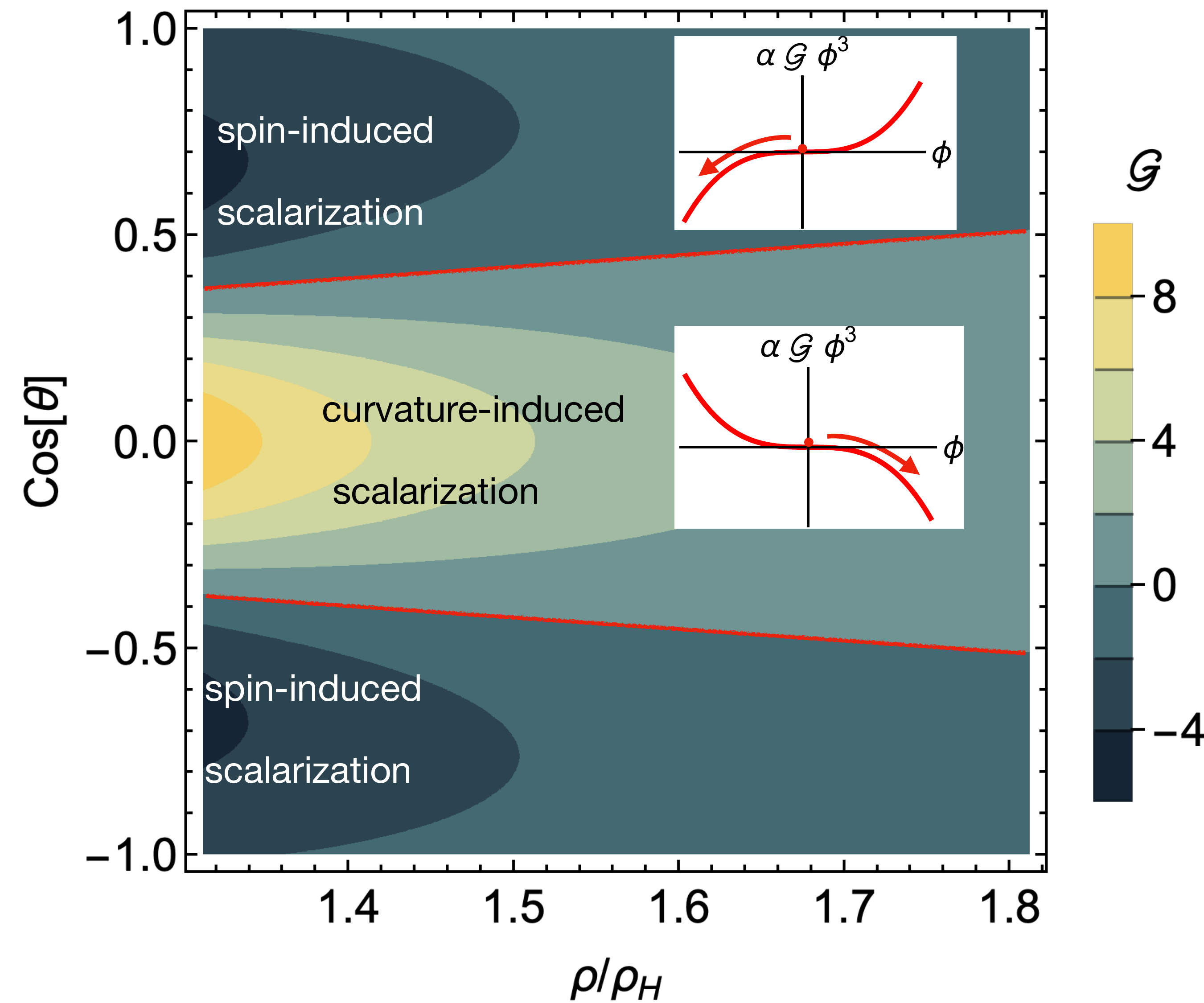}
    \caption{\label{fig:schematicscalarization}We plot the Gauss-Bonnet invariant (in units of the black-hole mass) as a function of $\rho$ and $\cos(\theta)$ for $j=0.95$. For a cubic interaction term and fixed sign of the coupling, e.g., $\alpha>0$, there is an unstable direction for the scalar field in all spacetime regions; those with positive Gauss-Bonnet (yellow tones) and those with negative Gauss-Bonnet (blue-gray tones). }
    \label{fig:GB_high_j}
\end{figure}

\begin{enumerate}
    \item In the regions where $\mathcal{G} > 0$, the no-hair theorem is violated if
    \begin{equation}
        \begin{aligned}
            &\alpha < 0 \quad \mathrm{and}\quad  \phi < 0,\qquad \mathrm{or}\\&
            \alpha > 0  \quad \mathrm{and}\quad  \phi > 0,
        \end{aligned}
    \end{equation}
    at least in some region of the spacetime.
    If non-trivial solutions exist, they would be the equivalent of the standard curvature-induced solutions \cite{Doneva:2017bvd,Silva:2017uqg,Cunha:2019dwb} in our theory. Because regions with $\mathcal{G}>0$ exist for any value of the spin, this way of violating the no-hair theorem is always present.
    \item In the regions where $\mathcal{G}<0$, which 
    is
    present for sufficiently high black-hole spins, the no-hair theorem is also violated if
    \begin{equation}
        \begin{aligned}
            &\alpha < 0 \quad \mathrm{and}\quad  \phi > 0,\qquad \mathrm{or}\\&
            \alpha > 0 \quad \mathrm{and}\quad  \phi < 0,
        \end{aligned}
    \end{equation}
    at least in some region of the spacetime.
    If such non-trivial solutions exist, they would be the equivalent of spin-induced solutions \cite{Dima:2020yac,Berti:2020kgk,Herdeiro:2020wei} in our theory. Because these regions occur in addition to the positive-$\mathcal{G}$ regions at high spin, $j>1/2$, black holes of high spin have two distinct ways of violating the no-hair theorem in our theory.
\end{enumerate}

We conclude that, regardless of the sign of the coupling constant $\alpha$, the no-hair theorem can, for the same theory, be violated in two qualitatively distinct ways, depending solely on the sign of the scalar field. For example, in Fig.~\ref{fig:GB_high_j}, we plot the profile of $\mathcal{G}$ in the region outside the horizon, where we observe regions with positive curvature and and regions negative curvature. In this case, the no-hair theorem could be broken in two distinct ways, as discussed above. This would mean that both curvature-induced and spin-induced solutions co-exist with the Kerr metric, at least at high enough spins, a feature which is not present in conventional scalarization models.
The co-existence of two distinct branches of scalarized solutions, each associated with a different breaking of the no-hair theorem, \emph{strongly} breaks black-hole uniqueness.  In such a setting, it is conceivable that the properties of a black hole of given mass and spin depend on its formation history.

\section{Methodology} \label{Methodology}
\subsection{Solving the equations of motion}
To solve the equations of motion, we adopt the numerical code developed in Ref.~\cite{Fernandes:2022gde}.
An axisymmetric, stationary, circular black-hole spacetime can be parameterized by the following line element written in quasi-isotropic coordinates $(t,r,\theta,\varphi)$~\cite{Xie:2021bur,Konoplya:2016jvv}:
\begin{equation} \label{eq:metric_num}
    \begin{aligned}
        ds^2 =& -f\Big(1-\frac{r_H}{r}\Big)^2 dt^2 + \frac{g}{f} \Bigg[ h (dr^2 + r^2 d\theta^2)\\&
        +r^2 \sin^2\theta \Big(d\varphi -\frac{W}{r^2}r_H dt\Big)^2 \Bigg],
    \end{aligned}
\end{equation}    
where $f$, $g$, $h$ and $W$ are undetermined functions of the radial coordinate $r$ and the polar angle $\theta$ and $r_H$ denotes the coordinate location of the horizon.
A solution to the equations of motion is found once the four functions $f(r,\theta)$, $g(r,\theta)$, $h(r,\theta)$ and $W(r,\theta)$ as well as the scalar field $\phi(r,\theta)$ are determined.

The modified Einstein field equations obtained from the action~\eqref{eq:action} read
\begin{equation} \label{EoM def}
   G_{\mu \nu} = 8\pi \, T_{\mu \nu},
\end{equation}
where $G_{\mu \nu}$ is the Einstein tensor and $T_{\mu\nu}$ takes the form~\cite{Fernandes:2022gde}
\begin{equation}
    T_{\mu \nu} = \nabla_\mu \phi \, \nabla_\nu \phi - \frac{1}{2} g_{\mu \nu} (\nabla \phi)^2 + \frac{\alpha}{6} {}^{\ast}R^{\ast}_{\mu \alpha \nu \beta} \nabla^{\alpha} \nabla^{\beta}\phi^3.
\end{equation}
Here, \( {}^{\ast}R^{\ast}_{\mu \alpha \nu \beta} \) is the double-dual Riemann tensor, defined as
\begin{equation}
    {}^{\ast}R^{\ast}_{\mu \alpha \nu \beta} = \frac{1}{4} \epsilon_{\alpha \beta \gamma \delta} R^{\rho \sigma \gamma \delta} \epsilon_{\rho \sigma \mu \nu}.
\end{equation}
The scalar field equation of motion is given in Eq.~\eqref{eq:EoM scalar field}.

This system of equations results in five linearly independent partial differential equations which we solve numerically using a pseudospectral method. 
We use the Chebyshev polynomials $T_i(x)$ as basis functions for the radial part, and even cosines for the angular part. To do so, we introduce a compactified radial coordinate
\begin{equation} \label{from x to r}
    x = 1- \frac{2\,r_H}{r}.
\end{equation}
With this choice of coordinates, the event horizon is located at $x=-1$ and spatial infinity corresponds to $x=+1$. Therefore, the radial coordinate $x$ spans the interval $x \in [-1,+1]$, which coincides with the natural domain of the Chebyshev polynomials.

We expand the functions in terms of the basis functions
\begin{equation} \label{num_approx}
     f(x, \theta) \simeq \frac{1}{2}\beta_0 + \sum_{i=1}^{N_x-1} \sum_{j=1}^{N_{\theta}-1} \beta_{ij} T_i(x) \cos (2j\theta),
\end{equation}
and similarly for the other four functions.
The parameters $N_x$ and $N_{\theta}$ denote the resolutions of the approximation in each coordinate direction, and the $\{ \beta_{ij} \}$ are the spectral coefficients which are derived using the Newton-Raphson method for root-finding, see Ref.~\cite{Fernandes:2022gde} for further details on the numerical method and convergence properties of the code.

\subsection{Physical properties of the black-hole solutions}
We study the physical properties of the black-hole solutions by calculating the ADM mass $M$, the angular momentum $J$, as well as mechanic and thermodynamic quantities such as the horizon area $A_H$, Hawking temperature $T_H$ and entropy $S$.
These are given by \cite{Fernandes:2022gde}
\begin{equation}
M = r_H (1+\partial_x f)|_{x=1},
\end{equation}
\begin{equation}
    J = -r_H^2 \partial_x W |_{x=1},
\end{equation}
\begin{equation}
T_H = \frac{1}{2\pi r_H} \frac{f}{\sqrt{gh}}\Bigg|_{x=-1},
\end{equation}
\begin{equation}
A_H = 2\pi _H^2 \int_0^{\pi} \dd \theta \sin \theta \frac{g\sqrt{h}}{f}\Bigg|_{x=-1}.
\end{equation}
The entropy is obtained using the Iyer-Wald formalism \cite{PhysRevD.50.846}
\begin{equation}
S = -2\pi _H^2 \int_H \dd A \frac{\delta \mathcal{L}}{\delta R_{\mu \nu \alpha \beta}}\epsilon_{\mu \nu} \epsilon_{\alpha \beta} 
\Bigg|_{\text{on-shell}},
\end{equation}
where $\mathcal{L}$ is the Lagrangian of the theory and $\epsilon_{\mu \nu}$ is the binormal vector to the event horizon surface.

Besides characterizing the black-hole spacetime, these quantities enable us to validate the accuracy of our code, by checking that the Smarr relation is satisfied. This relation reads~\cite{Fernandes:2022gde}
\begin{equation} \label{eq: smarr}
M = 2T_H S + 2 \Omega_H J - 2 \int_{\Sigma} \dd^3 x \sqrt{-g} \mathcal{L}\Bigg|_{\text{on-shell}},
\end{equation}
where
\begin{equation}
    \Omega_H = \frac{W}{r_H}\Bigg|_{x=1},
\end{equation}
is the angular velocity of the horizon.

Lastly, we define the scalar charge $Q_s$, which appears in the asymptotic expansion of the scalar field
\begin{equation}
    \phi = \frac{Q_s}{r}+ \mathcal{O}(r^{-2}).
\end{equation}
This allows us to distinguish the Kerr metric ($Q_s = 0$) from a hairy solution ($Q_s \neq 0$).\\

\section{Results}\label{Results}
In what follows, we fix $\alpha<0$, without loss of generality, since the theory is symmetric under the simultaneous mapping $\alpha\to-\alpha$ and $\phi\to -\phi$. According to the discussion in Sec.~\ref{Our theory}, we expect curvature-induced solutions with a negative scalar (at least in some spacetime regions) to exist across all values of the spin and for $M$ beyond a (spin-dependent) threshold. In addition, we expect spin-induced solutions with positive scalar (at least in some spacetime regions) beyond a critical value of the spin.

\subsection{Curvature-induced black holes} \label{sec:curvature induced section}
\begin{figure*}[!t]
\includegraphics[width=0.49\linewidth]{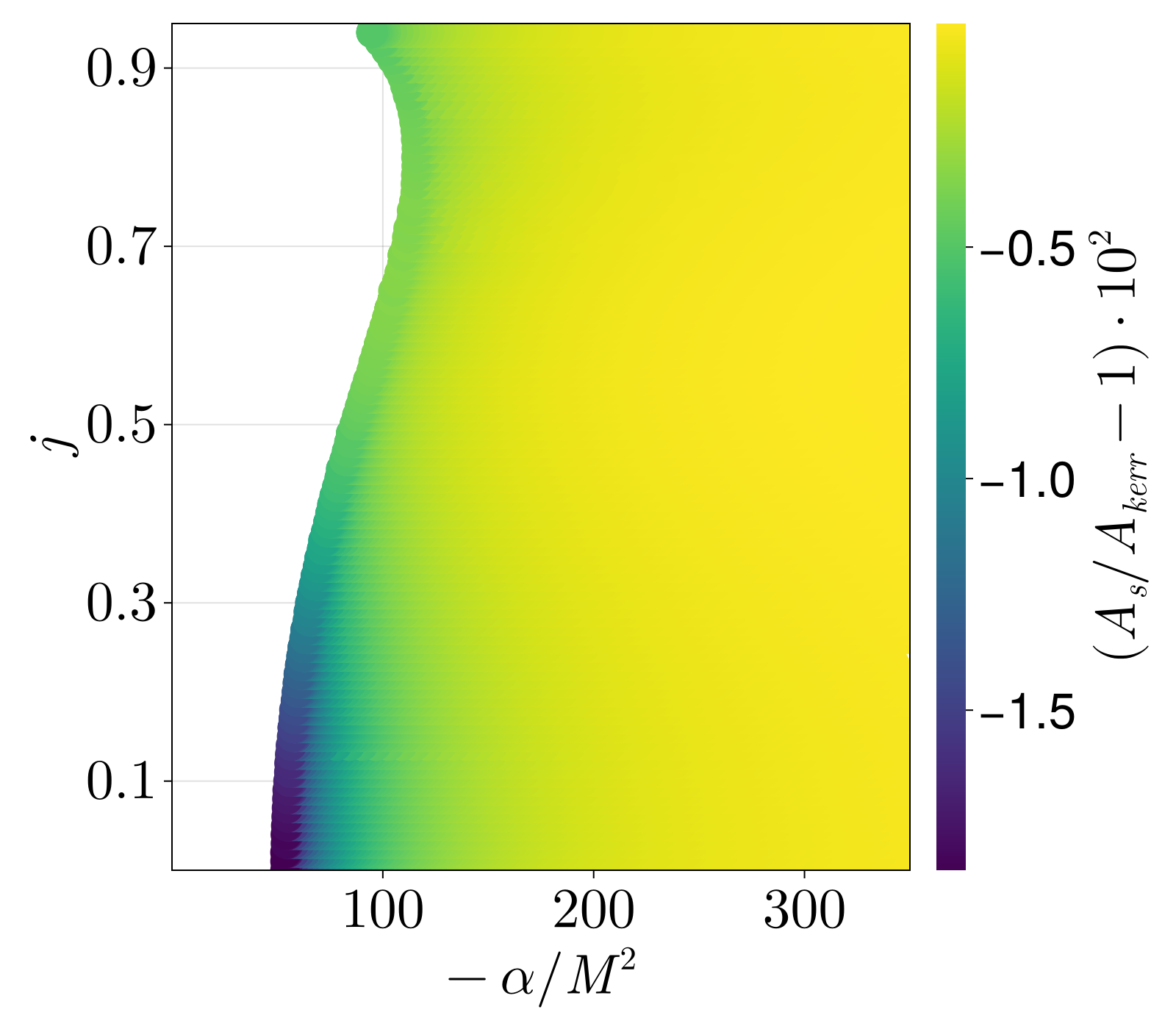}\quad
\includegraphics[width=0.49\linewidth]{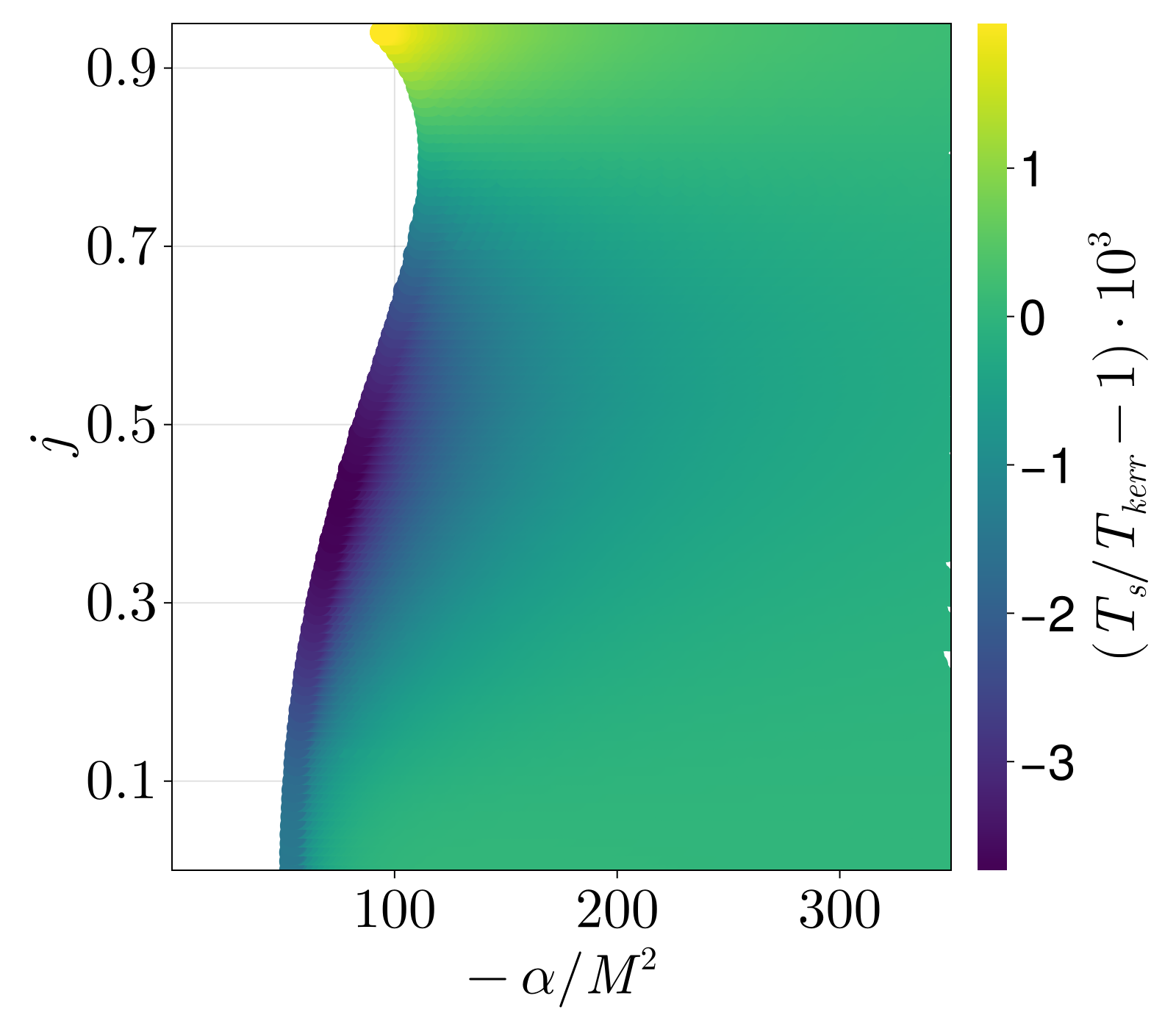} \\
\includegraphics[width=0.49\linewidth]{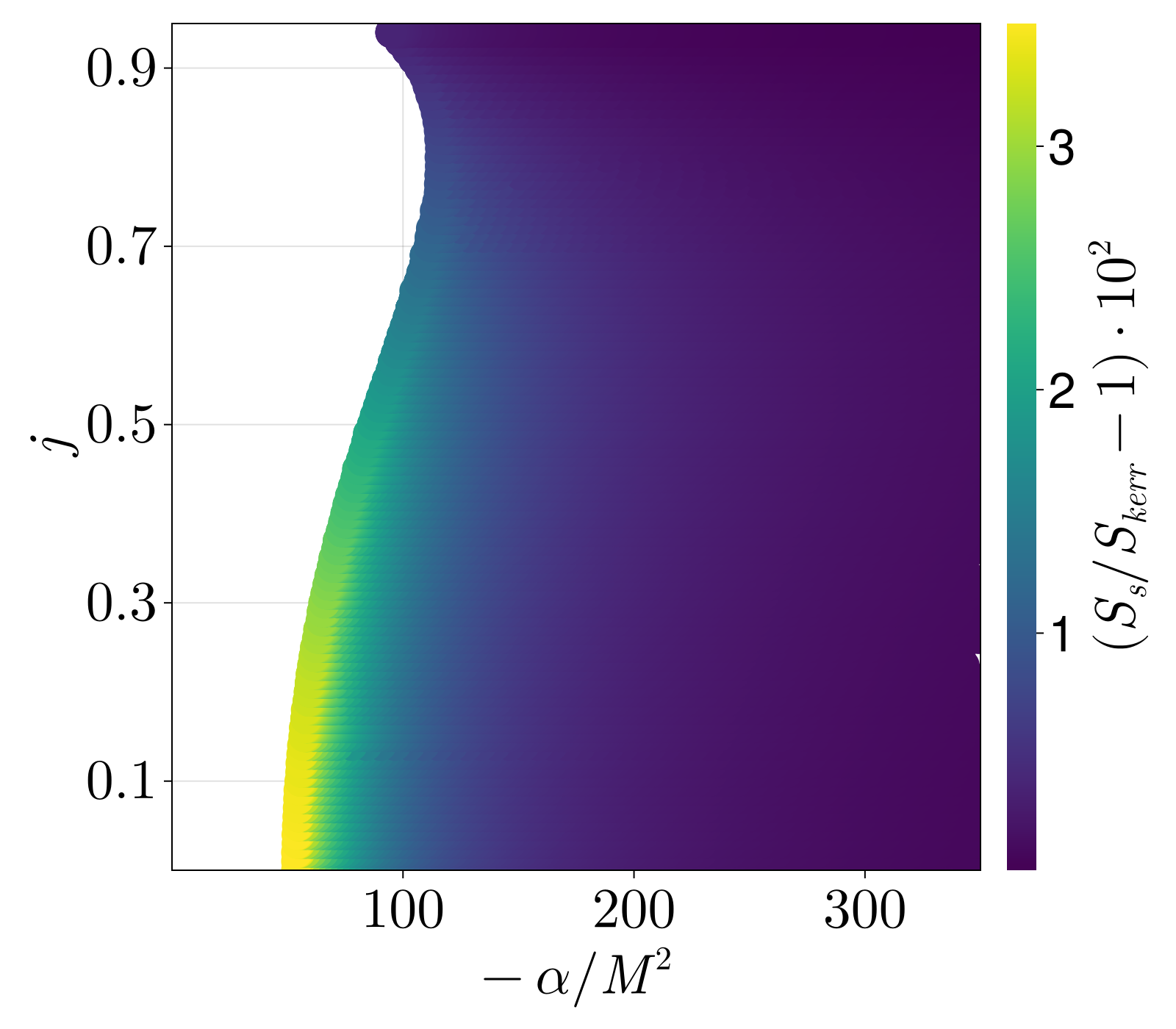}\quad
\includegraphics[width=0.49\linewidth]{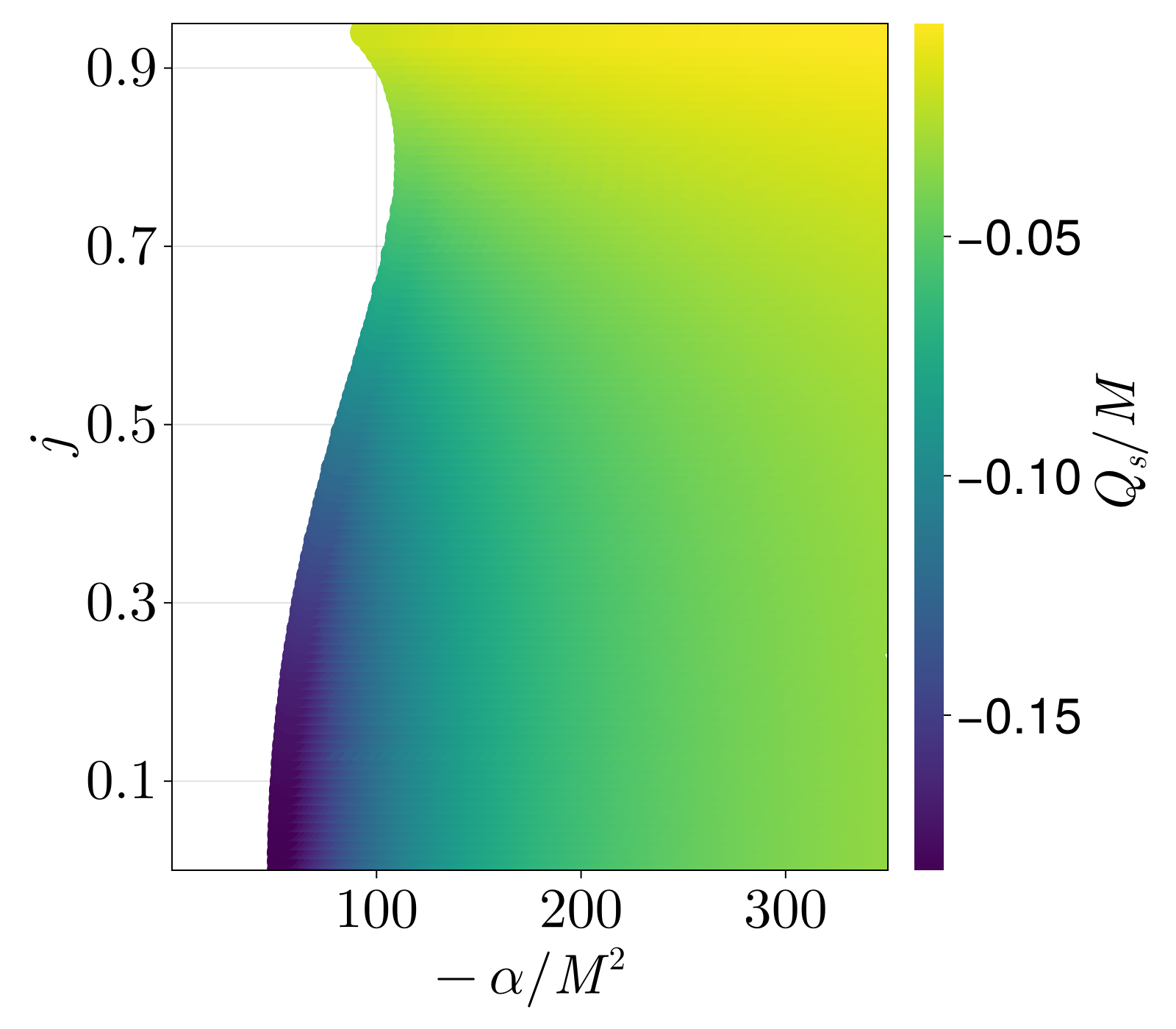} \\
    \caption{Domain of existence for curvature-induced solutions. The color bars display the deviations of the scalarized solutions (denoted by the subscript “s”) relative to the Kerr solution (denoted by the subscript “Kerr”). Each point in the domain of existence corresponds to a solution characterized by a dimensionless spin $j$ and a coupling parameter $-\alpha/M^2$. From top left to bottom right, we present the deviations in: the horizon area of the scalarized black hole, $A_s$; its Hawking temperature, $T_s$; its entropy, $S_s$; and the scalar charge of the field, $Q_s$.}
    \label{fig:numerical_solution_curvature}
\end{figure*}

We identify curvature-induced solutions by checking that the numerically obtained solution i) has a nonzero scalar charge, and is therefore not the Kerr solution; ii) its sign aligns with the expected negative sign; and iii) confirm that we can follow solutions continuously across the two-dimensional plane spanned by $-\alpha/M^2$ and $j$ starting from small $j$.
Our results are validated using the Smarr relation \eqref{eq: smarr}, which typically holds to a relative precision of  $\sim 10^{-7}$ or lower.

The domain of existence\footnote{We stopped generating solutions for $j > 0.94$ due to numerical limitations.} of curvature-induced solutions is plotted in Fig.~\ref{fig:numerical_solution_curvature}. 
The shape of its left boundary (at small $|-\alpha/M^2|$) is similar to that of the standard curvature-induced scalarized solutions with a $\phi^2$ coupling, see, e.g., Ref.~\cite{Fernandes:2022kvg}. However, in our case there is no maximum value of $-\alpha/M^2$ where scalarized solutions cease to exist; instead the Kerr solution is  approached continuously as $-\alpha/M^2\to +\infty$. As shown in Ref.~\cite{Fernandes:2022kvg}, the presence or absence of such a maximal value of $-\alpha/M^2$ depends sensitively on the specific coupling function of the theory. The $\phi^3$ coupling appears to belong to a class of models in which no upper bound on $-\alpha/M^2$ arises for the existence of scalarized solutions; see Ref.~\cite{Fernandes:2022kvg} for additional examples.

Deviations in geometrical quantities, such as $A_H$ are 
generally
small, of order $\mathcal{O}(10^{-2})-\mathcal{O}(10^{-3})$. The same is true for the location of the light ring and ISCO, which we do not explicitly show, because the relative deviations are small. This is a significant difference to scalarization through a tachyonic instability, which tends to lead to more sizeable deviations from the Kerr spacetime. It may also imply that scalarization through a non-linear instability is more difficult to constrain observationally.
However, the scalar charge achieves values $\mathcal{O}(10^{-1})$ in units of the black-hole mass.

From Fig.~\ref{fig:numerical_solution_curvature} we infer that moving from right to left along the $-\alpha/M^2$ axis, corresponding to the scalarized black hole 
increasing its mass, leads to a discontinuous transition, akin to a first-order phase transition in statistical physics, from the scalarized black-hole branch to the Kerr solution. This behavior is clearly seen in the scalar charge which, for small spins, drops discontinuously from values $Q_s/M \sim \mathcal{O}(10^{-1})$ to zero. We hypothesize that this may even be dynamically relevant, e.g., in a process where a black hole increases its mass by accretion.
By contrast, as $-\alpha/M^2$ increases further from within the scalarized regime, all the quantities we analyze smoothly converge to their corresponding values for the Kerr solution.

The thermodynamic properties of the curvature-induced solution may provide insight into its stability. A given solution is thermodynamically preferred over another if it possesses higher entropy. This, however, does not by itself imply a dynamical instability of the lower-entropy solution, although in scalarization scenarios the two notions often appear to be correlated; see, for example, Refs.~\cite{Herdeiro:2018wub,Fernandes:2019rez,Fernandes:2019kmh}\footnote{See, however, Ref.~\cite{Held:2022abx} for a counterexample in higher gravity.}. Moreover, higher-order scalar-Gauss-Bonnet interactions or Ricci couplings~\cite{Thaalba:2023fmq, Doneva:2024ntw, Ventagli:2021ubn, Fernandes:2024ztk} impact the endpoint of the non-linear instability even though they do not contribute to its onset. These couplings are expected to contribute to the entropy. We leave such considerations for future work.
In our case, the entropy of the curvature-induced scalarized solutions is always greater than that of a Kerr black hole, cf.~the lower left panel in Fig.~\ref{fig:numerical_solution_curvature}. With the above caveats in mind, we take this as a hint that the scalarized solution may be dynamically preferred, even though the dynamical stability of the solution is of course an independent question.

\subsection{Spin-induced black holes} \label{spin induced section}
\begin{figure*}[t!]
    \centering
    \includegraphics[width=0.49\linewidth]{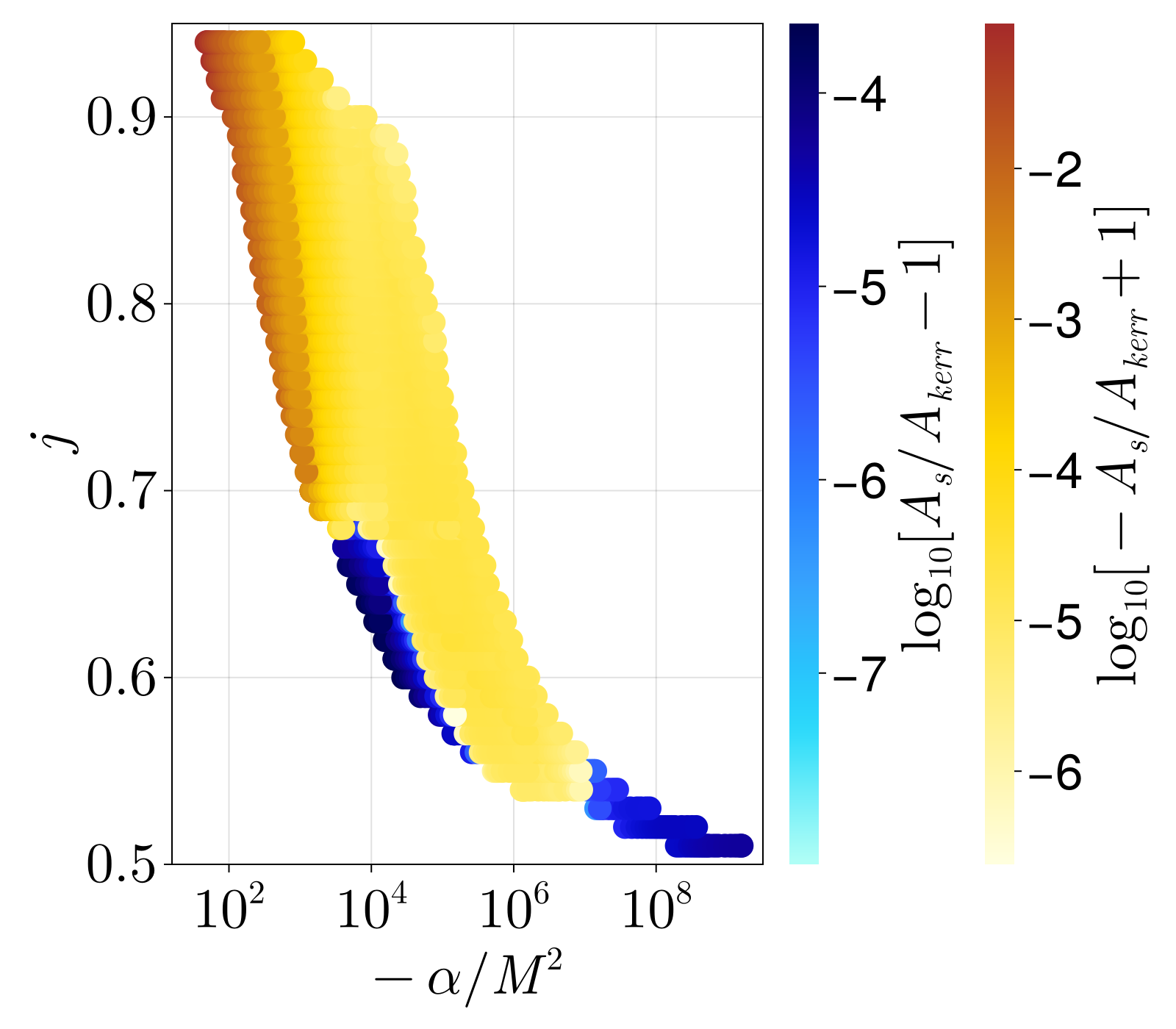}\quad
\includegraphics[width=0.49\linewidth]{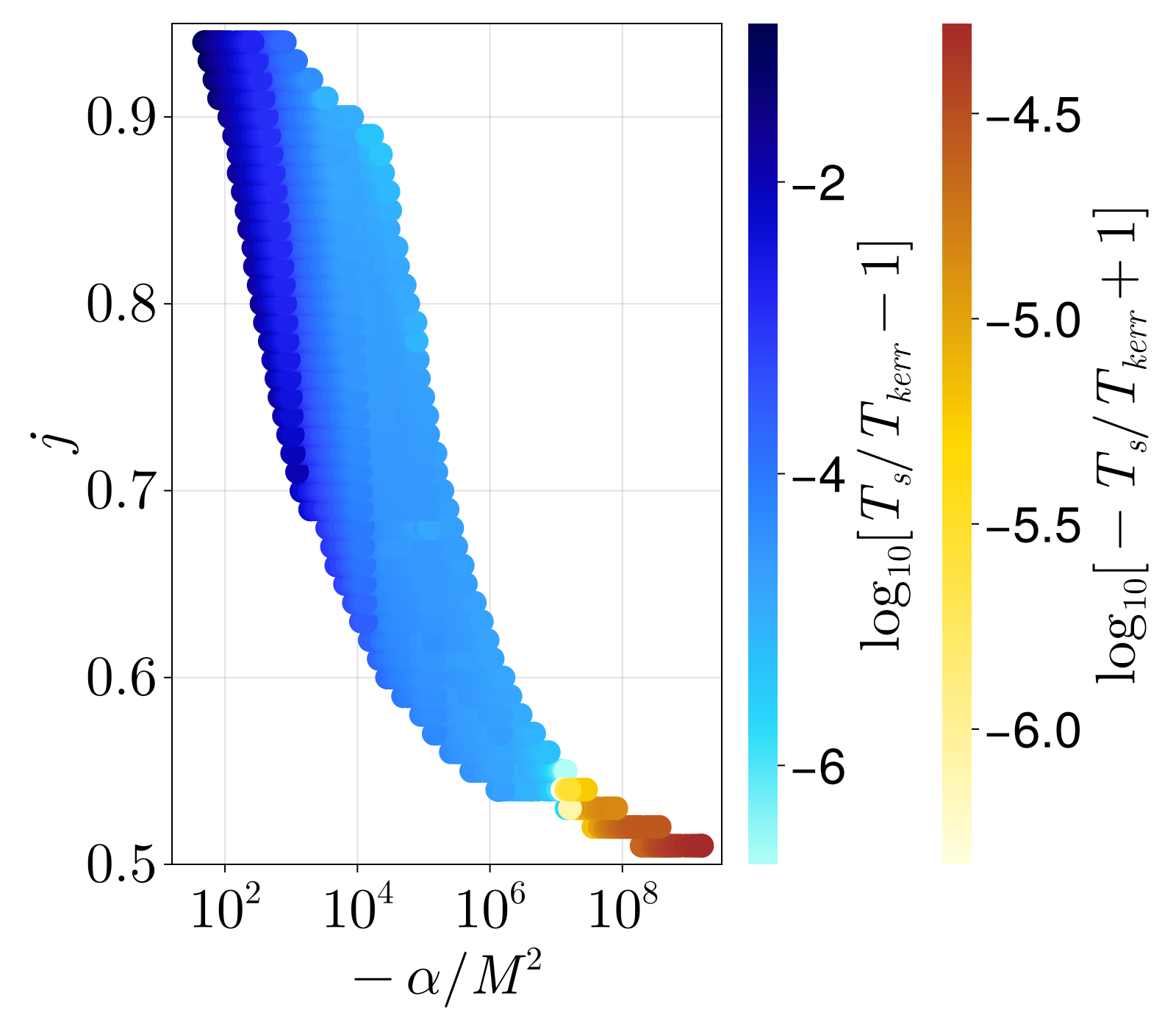}\\
\includegraphics[width=0.49\linewidth]{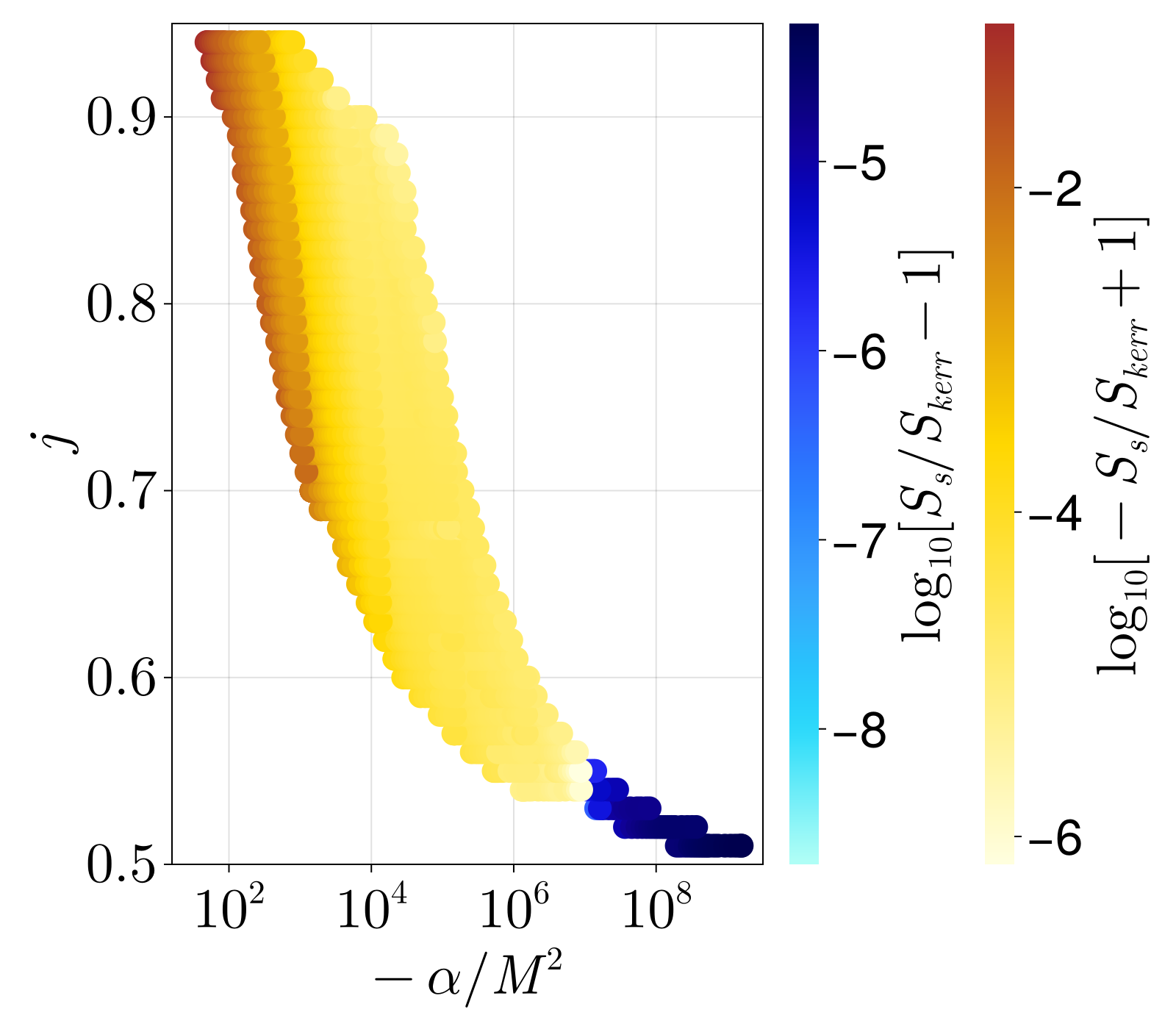}\quad
    \includegraphics[width=0.49\linewidth]{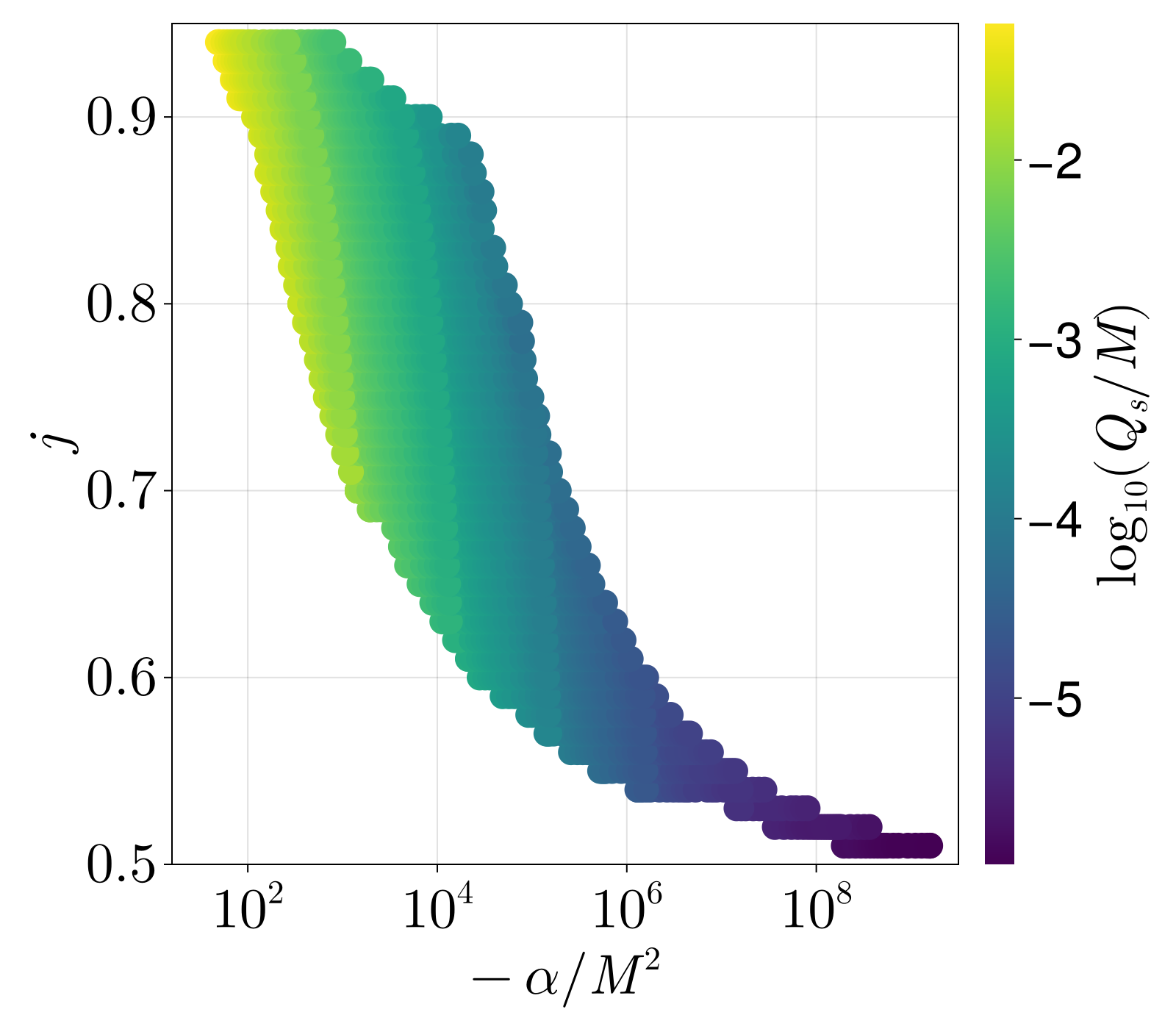}
    \caption{Domain of existence of spin-induced solutions. The color scale shows the deviation of the scalarized black hole (denoted by the subscript “s”) from the Kerr solution (denoted by the subscript “Kerr”). Each point corresponds to a solution specified by a coupling $-\alpha/M^2$ and a dimensionless spin $j$. From top left to bottom right, we display the deviations in: the horizon area of the scalarized black hole, $A_s$; its Hawking temperature, $T_s$; its entropy, $S_s$; and the scalar charge, $Q_s$.}
    \label{fig:numerical_solution_spin}
\end{figure*}

As before, we identify spin-induced solutions by the non-vanishing and positive value of the scalar charge, and by continuity arguments. Additionally, their region of existence is bounded from below by $j\approx 1/2$. The Smarr relation \eqref{eq: smarr} holds to a relative precision of  $\sim 10^{-6}$ or smaller. The domain of existence of spin-induced scalarized solutions is presented in Fig.~\ref{fig:numerical_solution_spin}, where the quantities $A_H$, $T_H$, $S$ and $Q_s$ are shown in the color bars.

As expected, the scalar charge is positive and generally grows with increasing spin. However, in contrast to the curvature-induced solutions, the entropy of the spin-induced solutions is only greater than that of a Kerr black hole with the same mass and spin for spins slightly above the threshold of existence, $j\approx 1/2$. A non-monotonic behavior of the difference of the entropy of a scalarized solution to a Kerr black hole has previously been observed in another scalar-Gauss-Bonnet theory \cite{Eichhorn:2023iab}. We leave the very interesting question of the dynamical implications of this behavior for future work.

Importantly, for spins above $j\approx1/2$, we find regions of the domain of existence where the three distinct branches of solutions (Kerr, curvature-induced and spin-induced solutions) co-exist for the same values of mass and spin. These regions are shown in Fig.~\ref{fig:phase diagram}. Within these regions, it may depend on the formation history of a black hole which branch it ends up with.

Because the scalar charge has the opposite sign relative to that of curvature-induced solutions, the various transitions between curvature-induced and spin-induced solutions are always discontinuous. Similarly, transitions from the Kerr solution to the spin-induced solution are always discontinuous, cf.~Fig.~\ref{fig:numerical_solution_spin}.

Once again, the absolute deviations from the Kerr solution are rather small. Although we have computed the deviations for the light ring and the ISCO, we do not display them here, as they are of order $\mathcal{O}(10^{-3})$-$\mathcal{O}(10^{-4})$. The largest deviations arise at high spin, where numerical difficulties prevent us from fully exploring the parameter space. The tendency for deviations to grow with increasing spin is also observed in spin-induced scalarization triggered by a tachyonic instability (see, e.g., Fig.~1 in \cite{Fernandes:2024ztk}); in that case, however, the absolute deviations are significantly larger.

In terms of observational tests of strong breaking of black-hole uniqueness, our results thus suggest the following picture: for the stationary solutions, a detection of the deviation from the Kerr solution may be challenging, given the relative smallness of the deviations. In contrast, for time-dependent settings, there may be qualitatively distinct effects in our theory. Specifically, discontinuous transitions between the Kerr solution and a scalarized solution, triggered, e.g., by accretion of mass, may result in observational signatures, e.g., through gravitational-wave emission. This suggests that an assessment of the detectability of strong breaking of black-hole uniqueness may rely on going beyond the analysis of stationary solutions.

\section{Discussion} \label{Conclusions}
\begin{figure*}[t!]
    \centering
    \includegraphics[width=0.65\linewidth]{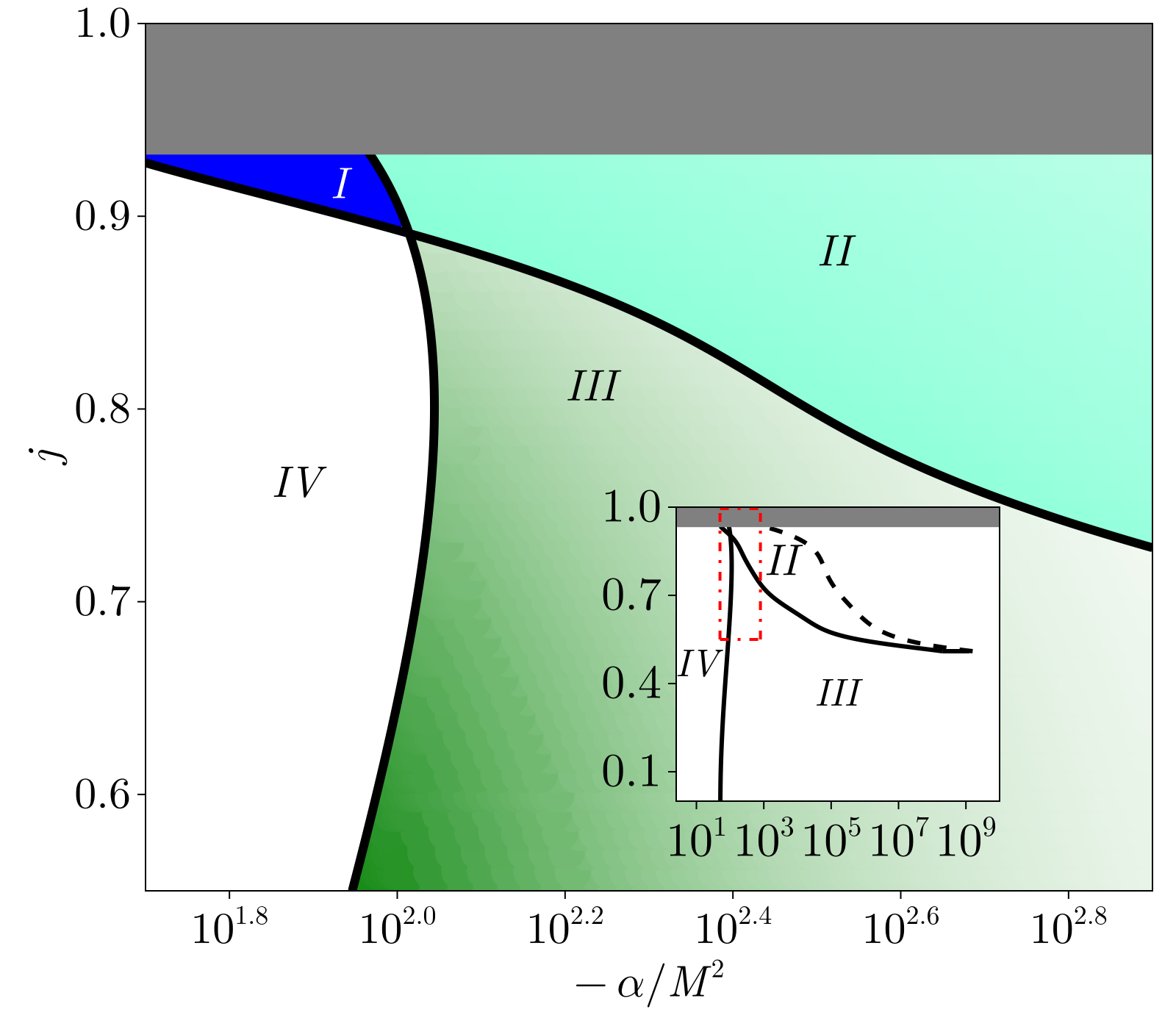}
    \caption{Phase diagram of the theory. The plot is split in four regions: in region \textit{I}, Kerr and spin-induced solutions co-exist; in region \textit{II} Kerr, spin- and curvature-induced solutions co-exist. This is the region where black-hole uniqueness is strongly broken; in region \textit{III} Kerr and curvature-induced solutions co-exist and in region \textit{IV} only the Kerr solution is present. The main plot is a zoom-in of the whole phase diagram, presented in the inset. The region of the whole phase diagram we zoom in is highlighted with a blue box in the inset. A continuous line describes a discontinuous phase transition and a dashed line a continuous one.}
    \label{fig:phase diagram}
\end{figure*}

In this work, we have explored black-hole scalarization which is triggered by a non-linear, rather than a tachyonic instability. We confirm that, as one may expect, a non-linear instability, encoded in a $\phi^3\, \mathcal{G}$-coupling, is effective at triggering scalarization.

Theories in which black holes scalarize violate no-hair theorems that hold in GR. Thereby they also violate black-hole uniqueness theorems, because the Kerr geometry remains a solution of the theory, albeit it is an unstable one in  part of the parameter space. Tachyonic scalarization mechanisms, do, however, only give rise to one new branch of solutions for any given value of the coupling, either curvature-induced scalarization (for one sign of the coupling) or spin-induced scalarization (for the other sign of the coupling).

For the scalar-Gauss-Bonnet theory with a cubic coupling that we study here, black-hole uniqueness is violated more strongly, if we quantify the amount of violation by the number of branches of beyond-GR-solutions. For large enough spin, and large enough value of the coupling, curvature-induced \emph{and} spin-induced scalarization mechanisms compete. In such a theory, the question, which black-hole geometry is realized may, e.g., depend on the formation history of the black hole.

Such a ``strong" violation of black-hole uniqueness may not only be interesting for black-hole observations, but also gives rise to an interesting \emph{phase diagram}: Spanning the phase diagram for black-hole solutions by $-\alpha/M^2$ and $j$, we find several transition lines:
The $K-C$-transition line describes the transition between the Kerr geometry and curvature-induced scalarization. Across this transition, $Q_s$, horizon-area, entropy and temperature jump. If we interpret one of these quantities as an order parameter for the transition, we identify the transition as a discontinuous one, akin to first-order transitions in statistical physics. 

The $K-S$-transition line describes the transition between the Kerr solution and spin-induced solution. It is again a discontinuous transition.
Finally, the $S-C$-transition line describes the transition from spin-induced to curvature induced, which is also discontinuous, because the sign of the scalar charge changes.

The line we expect to be a continuous transition line, i.e., akin to phase transitions of higher order in statistical physics, is the $C-K$ line at very large values of the coupling.
In Figs.~\ref{fig:phase diagram}, we observe how the quantities characterizing the solution trend back towards their Kerr values, motivating our expectation of a higher-order line. Numerically, we can of course not distinguish whether this is a transition at finite, but large values of $-\alpha/M^2$, or whether the Kerr solution is approached asymptotically, for $-\alpha/M^2 \rightarrow \infty$.\\ 
This phase diagram clearly motivates further study, not least because one may expect that dynamical transitions across these transition lines, were they realized, e.g., in the evolution history of a given black hole, may be probed through observations. They may, e.g.,  be accompanied by the emission of gravitational and/or scalar waves, potentially resulting in observational signatures.

Let us also compare our phase diagram to the corresponding phase diagram for scalar-Gauss-Bonnet theories with a tachyonic instability. There, the $C-S$- and $S-C$- transition lines are missing; instead, one must draw the phase diagram across both signs of the scalar-Gauss-Bonnet coupling. Then, there is a $K-S$- and $S-K$- transition line, the second of which appears to describe a continuous transition rather than a discontinuous one
(cf.~Fig.~1 in \cite{Fernandes:2024ztk}).\\ 

Our discussion so far has ignored the question whether scalarized black holes are \emph{stable} solutions of the theory.
To start addressing this question, we have investigated radial stability of the static scalarized solutions and found evidence for instability.\footnote{We study the integral of the effective potential that governs the radial perturbation equations. A negative value of this integral implies instability \cite{Blazquez-Salcedo:2018jnn}.} We expect this to extend at least to the curvature-induced solutions at low spin, because there is no reason to expect only the static limit to be unstable.

It is known that a non-minimal coupling between the scalar and the Ricci scalar can mitigate instabilities~\cite{Thaalba:2023fmq,Ventagli:2021ubn,Doneva:2024ntw,Fernandes:2024ztk,Thaalba:2024htc}. Following these works, we have done a preliminary study of the effects of adding a Ricci coupling in our theory. We find indications that the Ricci coupling may stabilize the static black holes and expect that this extends at least to low spins for stationary solutions. 

As a second pathway to stability, one may expect that higher-order (e.g., quartic or higher) terms in the coupling may stabilize the theory. We have again found indications that this is indeed the case.\footnote{We do so by again studying the integral of the effective potential. A positive value is not sufficient for stability; however, a negative value is sufficient for instability. We find a positive value, thus showing that a sufficient condition for instability is not realized.} 
We have not studied the stability of spin-induced solutions and leave this challenging question for future work.

Our work is a starting point for several distinct lines of research. First, understanding whether any of the scalarized solutions are stable in their domain of existence is highly relevant. It is also particularly interesting to understand whether curvature-induced and spin-induced scalarization both provide competing scalarization channels whose endpoint is a stable black hole for the same values of mass and spin. If so, our theory constitutes an example in which the formation history of a black hole determines which of several viable black-hole branches it belongs to.\\
Second, understanding observational signatures, e.g., in gravitational-wave signals, is of interest. This pertains in particular to the novel possibilities our theory opens up, where, e.g., at high spin, scalarized black holes might belong to two different branches, leading to potentially new imprints in gravitational-wave signals.\\
Third, neutron stars usually impose some of the strongest observational constraints on scalar-Gauss-Bonnet-theories~\cite{Danchev:2021tew}. This motivates their study within the present framework.\\
Fourth, the phase diagram we have presented in Fig.~\ref{fig:phase diagram} is intriguing not only because of the potential observational signatures of various transitions, but also because of its richness and in particular the point at which several transition lines intersect. Elaborating on the comparison to phase diagrams in statistical physics by defining suitable order parameters for the transitions and constructing effective potentials for them is of obvious interest.\\
Fifth, while scalar-Gauss-Bonnet theories are often motivated from string theory~\cite{Ferrara:1996hh,Antoniadis:1997eg,1985PhLB..156..315Z,Nepomechie:1985us,Callan:1986jb,Candelas:1985en,Gross:1986mw}, a thorough investigation into its status as a fundamental theory is called for. In particular, it would be relevant to understand whether scalar-Gauss-Bonnet theory can consistently be embedded not just into string theory, but also other quantum theories of gravity. 
A corresponding work is forthcoming \cite{deBrito2026}. In addition, different coupling functions $f(\phi)$ may arise in distinct quantum theories of gravity; or some coupling functions may lie in the absolute swampland \cite{Eichhorn:2024rkc}, i.e., not be compatible with any UV completion in quantum gravity at all.\newline \\

\section*{Acknowledgments}
We thank Leonardo Gualtieri for an interesting discussion on black-hole solutions in cubic scalar-Gauss-Bonnet theories.
We acknowledge the European Research Council's (ERC) support under the European Union’s Horizon 2020 research and innovation program Grant agreement No.~101170215 (ProbeQG).
This work is partially funded by the Deutsche Forschungsgemeinschaft (DFG, German Research Foundation) under Germany’s Excellence Strategy EXC 2181/1 - 390900948 (the Heidelberg STRUCTURES Excellence Cluster).

\bibliographystyle{apsrev4-2}
\bibliography{References}

\end{document}